\documentclass[10pt,letterpaper]{article}
\usepackage[top=0.85in,left=2.75in,footskip=0.75in]{geometry}

\usepackage{amsmath,amssymb}

\usepackage{changepage}

\usepackage[utf8x]{inputenc}

\usepackage{textcomp,marvosym}

\usepackage{cite}

\usepackage{nameref,hyperref}

\usepackage[right]{lineno}

\usepackage{microtype}
\DisableLigatures[f]{encoding = *, family = * }

\usepackage[table]{xcolor}

\usepackage{array}

\newcolumntype{+}{!{\vrule width 2pt}}

\newlength\savedwidth

\raggedright
\usepackage[aboveskip=1pt,labelfont=bf,labelsep=period,justification=raggedright,singlelinecheck=off]{caption}

\makeatletter
\renewcommand{\@biblabel}[1]{\quad#1.}
\makeatother

\date{}

\usepackage{lastpage,fancyhdr,graphicx}
\usepackage{epstopdf}
\fancyheadoffset[L]{2.25in}
\fancyfootoffset[L]{2.25in}
\usepackage{bm}

\begin{document}
\vspace*{0.2in}

\begin{flushleft}
{\Large
\textbf\newline{Stronger selection can slow down evolution driven by recombination on a smooth fitness landscape} 
}
\newline
\\
Masahiko Ueda\textsuperscript{1*},
Nobuto Takeuchi\textsuperscript{2},
Kunihiko Kaneko\textsuperscript{1,2}
\\
\bigskip
\textbf{1} Department of Basic Science, The University of Tokyo, Komaba, Meguro-ku, Tokyo 153-8902, Japan
\\
\textbf{2} Research Center for Complex Systems Biology, The University of Tokyo, Komaba, Meguro-ku, Tokyo 153-8902, Japan
\\
\bigskip

%
%





* ueda@complex.c.u-tokyo.ac.jp

\end{flushleft}
\section*{Abstract}
Stronger selection implies faster evolution---that is, the greater the
force, the faster the change. This apparently self-evident proposition,
however, is derived under the assumption that genetic variation within a
population is primarily supplied by mutation (i.e.\ mutation-driven
evolution). Here, we show that this proposition does not actually hold
for recombination-driven evolution, i.e.\ evolution in which genetic
variation is primarily created by recombination rather than mutation.
By numerically investigating population genetics models of
recombination, migration and selection, we demonstrate that stronger
selection can slow down evolution on a perfectly smooth fitness
landscape. Through simple analytical calculation, this apparently
counter-intuitive result is shown to stem from two opposing effects of
natural selection on the rate of evolution. On the one hand, natural
selection tends to increase the rate of evolution by increasing the
fixation probability of fitter genotypes. On the other hand, natural
selection tends to decrease the rate of evolution by decreasing the
chance of recombination between immigrants and resident individuals. As
a consequence of these opposing effects, there is a finite selection
pressure maximizing the rate of evolution. Hence, stronger selection can
imply slower evolution if genetic variation is primarily supplied by
recombination.



\section*{Introduction}
It is commonly expected that the rate of evolution is higher when
selection is stronger \cite{HarCla1997}.  This is because stronger
selection ensures fitter genotypes created by mutation to survive.
Indeed, it is well known that in a weak mutation regime (i.e.\ for
sufficiently low mutation rates) and on a smooth fitness landscape
\cite{TLK1996, KLRT1997}, the rate of evolution $v$ is described as
\begin{eqnarray}
 v &=& 4Nus,
 \label{eq:v_mutation}
\end{eqnarray}
where $N$ is the population size, $u$ is the beneficial mutation rate,
and $s$ is the selection coefficient \cite{Gil1998}. This equation shows
that the rate of evolution $v$ increases linearly with the strength of
selection $s$. Such monotonic dependence of $v$ on $s$ is expected to
persist even in a strong mutation regime. In this regime, beneficial
mutations can arise simultaneously and interfere with each other's
fixation, a phenomenon known as clonal interference. Although clonal
interference decreases $v$, making it less than proportional to $N$ and
$u$ (cf.\ Eq~\ref{eq:v_mutation}), $v$ still increases monotonically
with $s$ \cite{GerLen1998, Wil2004, DesFis2007, ParKru2007,
  GRBHD2012}. Therefore, there is no finite value of selection pressure
maximizing the speed of evolution.

However, this monotonic dependence of $v$ on $s$ has been derived under
the assumption that genetic variation within a population is primarily
created by mutation (mutation-driven evolution, for short). In this
paper, we show that this widely-known relationship does not actually
hold if genetic variation within a population is primarily created by
migration and recombination (recombination-driven evolution, for short),
even for a smooth fitness landscape. Recombination is a source of new
genotypes besides mutation. Recombination between genomes occurs in
sexual reproduction and is beneficial in avoiding Muller's ratchet and
clonal interference \cite{May1978, NSF2010, WWW2011}.  Furthermore, a
type of recombination known as horizontal gene transfer is considered to
be important also in the evolution of prokaryotes \cite{OLG2000,
  KMA2001, TKK2014, TCKK2015, NMFF2015, DPM2016}.

Specifically, we consider a situation in which novel genes are supplied
to a population through immigration from other populations followed by
recombination between migrant and resident individuals
(i.e. introgression). To ensure the generality of results, we
investigate two models representing distinct evolutionary scenarios. The
first model considers migration and recombination between populations
adapting to multiple distinct ecological niches (Model 1). The second
model considers migration and recombination between populations adapting
to a single common ecological niche (Model 2). Under both the models, we
find that there is an optimal selection pressure maximizing the
speed of evolution; i.e., $v$ is a non-monotonic function of $s$.

\section*{Model 1}
We assume that there are many populations, each of which is evolving
toward adaptation to a distinct ecological niche. A population
occasionally receives immigrants from the other populations. The
immigrants always have fitness lower than that of resident individuals
owing to differences in niches. However, the genomes of the immigrants are
assumed to contain genes that are beneficial to the resident
individuals, but are absent in the latter. These genes can be
transferred to the latter through recombination. For simplicity, only
the dynamics of a single population is explicitly considered, with that
of the others abstracted away on the basis of the mean-field-like
approximation as described below.

Throughout the paper, mutation is assumed to be rare enough to be
negligible in order to focus on recombination-driven evolution. The
fitness landscape is assumed to be smooth so that the fitness landscape
in itself does not cause the non-monotonic dependence of the speed of
evolution on selection pressure (see also Discussion).

\subsection*{Methods}

Model 1 assumes a population of $N$ individuals (see
Table~\ref{tbl:notation} for notation). The genotype of individual $i\in
\left\{ 1, \cdots, N \right\}$ is denoted by $\textrm{\boldmath $g$}_i
\equiv \left( g_{1,i}, \cdots, g_{L,i} \right)$. Each variable $g_{l,i}$
denotes a type of a gene (i.e. allele) at locus $l\in \left\{1, \cdots,
L\right\}$ and assumes either the value of $-1$ (deleterious) or $1$
(beneficial) \cite{Pel1997}.

\begin{table}
\caption{List of symbols.}
\begin{tabular}{|c|l|}
  \hline 
  $N$ & population size\\
  $L$ & the number of gene loci\\
  $r$ & recombination rate per individual per loci\\
  $\mu$ & migration rate for Model 1\\
  $D$ & migration rate for Model 2\\
  $N_\mathrm{s}$ & the number of subpopulations in Model 2\\
  $s$ & selection pressure\\
  $g_l$ & gene at locus $l$ ($g_l\in \{-1,1\}$) \\
  $\bm{g}$ & genotype defined as $\left( g_{1}, \cdots, g_{L}\right)$ \\
  $\phi(\bm{g})$ & rescaled fitness defined as $\sum_l g_l$ (fitness is
  defined as $e^{s \phi\left( \textrm{\boldmath $g$} \right)}$)\\
  $\phi_0$ & difference in rescaled fitness between resident individuals and migrants\\
  $v$ & the rate of evolution defined as $\left\langle \phi \right\rangle=vt+\mathrm{const.}$\\
  \hline 
\end{tabular}
\label{tbl:notation}
\end{table}

We consider the time evolution of the system, which consists of three
discrete steps: selection, recombination and migration.  In the
selection step, $N$ genotypes are selected from the present population
with probabilities proportional to the fitness of genotypes. The fitness
of genotype $\bm{g}$ is defined as $\exp\left( s \phi\left( \bm{g}
  \right) \right)$, where $\phi(\bm{g})$ is rescaled fitness defined as
\begin{eqnarray}
 \phi\left( \textrm{\boldmath $g$} \right) &\equiv& \sum_{l=1}^L g_l
 \label{eq:fitness}
\end{eqnarray}
($(\phi+L)/2$ counts the number of beneficial alleles in a genome), and
$s$ is selection pressure. Accordingly, the probability that individual
$j$ is selected for reproduction is
\begin{eqnarray}
 P\left( j \right) &=& \frac{e^{s\phi\left( \textrm{\boldmath $g$}_j \right)}}{\sum_{i=1}^N e^{s\phi\left( \textrm{\boldmath $g$}_i \right)}}.
 \label{eq:selection}
\end{eqnarray}
Note that if $\phi(\bm{g}_j) > \phi(\bm{g}_k)$, $P(j)/P(k)$ increases
with $s$; thus, the larger the value of $s$, the stronger natural
selection. Note also that the fitness landscape is smooth because it
contains only one local and global maximum and one local and global
minimum.

In the recombination step, individuals exchange genes with probability
$r$ per individual per locus per generation:
\begin{eqnarray}
 && \left( \cdots, g_{l-1,i}, g_{l,i}, g_{l+1,i}, \cdots \right) + \left( \cdots, g_{l-1,j}, g_{l,j}, g_{l+1,j}, \cdots \right) \nonumber \\
 &\rightarrow& \left( \cdots, g_{l-1,i}, g_{l,j}, g_{l+1,i}, \cdots \right) + \left( \cdots, g_{l-1,j}, g_{l,i}, g_{l+1,j}, \cdots \right).
 \label{eq:recombination}
\end{eqnarray}
Pairs of individuals undergoing recombination are selected randomly.

Migration occurs from the other populations (pool) to the system.  We
assume that individuals change with probability $\mu$ as
\begin{eqnarray}
 \textrm{\boldmath $g$}_i &\rightarrow& \textrm{\boldmath $g$}^{(\mathrm{pool})},
\end{eqnarray}
where genotype $\textrm{\boldmath $g$}^{(\mathrm{pool})}$ is randomly
generated with rescaled fitness $\phi\left( \textrm{\boldmath
    $g$}^{(\mathrm{pool})} \right) = \phi\left( \textrm{\boldmath $g$}_i
\right) - \phi_0$ with $\phi_0>0$.  That is, migration always decreases
the fitness of the system, but sequences $\textrm{\boldmath
  $g$}^{(\mathrm{pool})}$ and $\textrm{\boldmath $g$}_i$ are
uncorrelated (for this reason, the effect of migration differs from that
of introducing $\phi_0/2$ deleterious mutations).  The rescaled fitness
difference $\phi_0$ between a resident individual and a migrant is set
constant under the assumption that individuals in the other populations
also evolve at the same rate as those in the focal population.

For each simulation, the model was initialized with individuals having
random genotypes and rescaled fitness $\phi=0$. The parameters were set
as follows: $N=1000$ or $2000$, $L=1000$, $r=10^{-4}$ or $2\times
10^{-4}$, $\mu=10^{-3}$ or $2\times 10^{-3}$, and
$\phi_0=20$. Statistical quantities were calculated by running $1000$
replicate simulations.

\subsection*{Results}
\begin{figure}[!h]
\begin{adjustwidth}{-2.25in}{0in}
  \flushright\includegraphics[clip]{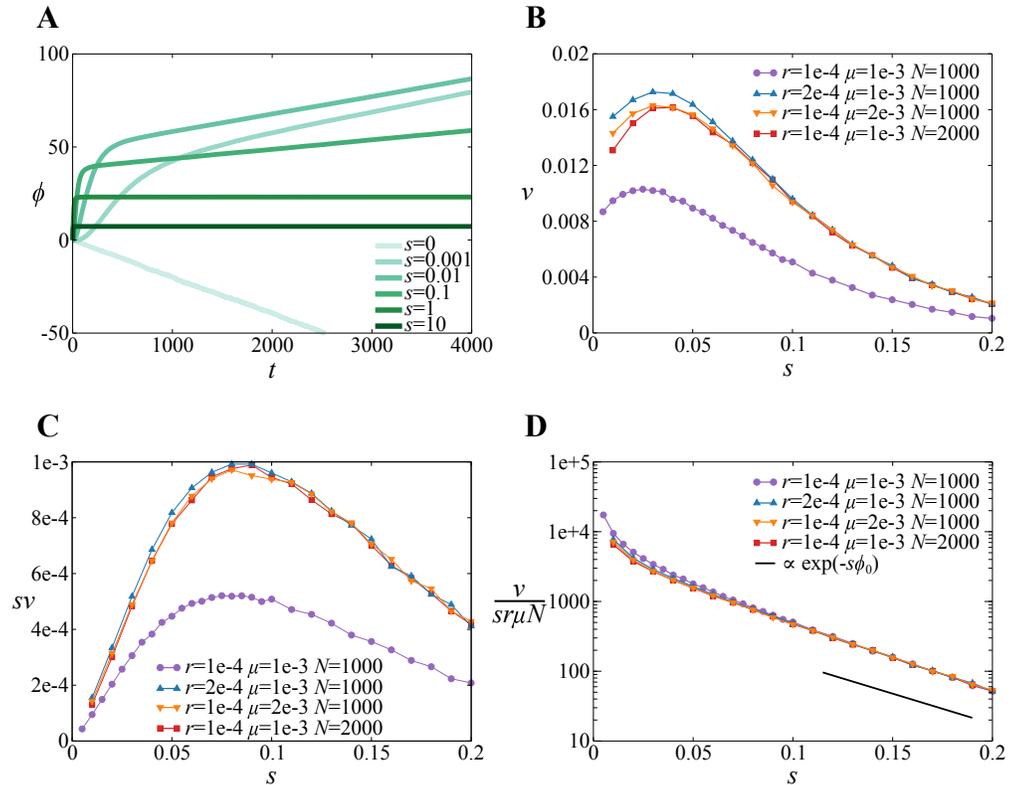}
  \caption{\textbf{A.} The average rescaled fitness $\left\langle \phi
    \right\rangle$ as a function of time $t$ for various strengths of
    selection (denoted by $s$) for Model 1. $\phi$ is proportional to
    the number of beneficial alleles in a genome (which is
    $(\phi+L)/2$). The parameters are as follows: $N=1000$, $L=1000$,
    $r=10^{-4}$, $\mu=10^{-3}$, and $\phi_0=20$ (see
    Table~\ref{tbl:notation} for notation). \textbf{B.}  The rate of
    evolution $\Delta \phi/\Delta t$ (denoted by $v$) as a function of
    $s$ for Model 1. $L=1000$ and $\phi_0=20$ (the other parameters are
    indicated in the graph). \textbf{C.} The rate of evolution in
    Malthusian fitness (i.e.\ a logarithm of fitness) $sv$ as a function
    of $s$ for Model 1. The parameters are the same as in B. \textbf{D.}
    A semi-log plot of $v/(sr\mu N)$. The parameters are the same as in
    B. The slope of the solid line is $-\phi_0$.}
\label{fig:model1}
\end{adjustwidth}
\end{figure}

We numerically calculated the time evolution of the average rescaled
fitness $\left\langle \phi \right\rangle$, where $\left\langle \cdots
\right\rangle$ denotes a population average. The result indicates that
the dynamics of $\left\langle \phi \right\rangle$ has two phases as
described below (Fig~\ref{fig:model1}A).

In the first phase, $\left\langle \phi \right\rangle$ rapidly increases
in a sigmoidal manner, except for $s=0$ (Fig~\ref{fig:model1}A, $t <
1/s$). During this phase, the effect of migration is negligible because
the migration rate $\mu$ is set to a value smaller than or equal to $s$
(specifically, $\mu=10^{-3}$). Thus, this phase constitutes an initial,
transient dynamics before migration takes effect, resulting from
selection and recombination within a population. During this phase, the
system becomes increasingly homogeneous as selection removes genetic
variations with virtually no supply of new genes through
migration. Eventually, one genotype is selected, whose fitness depends
on the selection pressure and the recombination rate. Since the
frequency of the fittest genotype increases exponentially as $e^{st}$,
the first phase lasts until $e^{st}\sim 1$. Therefore, the duration of
the first phase scales approximately as $1/s$.

In the second phase, $\left\langle \phi \right\rangle$ increases almost
linearly at a rate that depends on the value of $s$
(Fig~\ref{fig:model1}A, $t>1/s$). In this phase, a quasi-steady state is
achieved, in which a population is almost homogeneous, but continually
receives immigrants and incorporates new genes supplied by them through
recombination and selection. Thus, this phase constitutes evolution
driven by recombination and migration. Note, however, that $\left\langle
  \phi \right\rangle$ eventually saturates on an even longer time scale
($t\gg 4000$) as a trivial consequence of the fact that $\phi$ has the
maximum value $L$. Since we are interested in the rate of evolution
driven by recombination and migration, we hereafter focus on the second
phase of the dynamics well before this saturation occurs.

Note a special case arising for $s=0$, for which $\left\langle \phi
\right\rangle$ decreases monotonically (Fig~\ref{fig:model1}A). This
decrease is due to the assumption that immigrants always have fitness
lower than that of residents owing to differences in niches---the
assumption that becomes senseless when $s=0$. We do not consider this
special case hereafter because we are interested in the evolution under selective pressure and in testing the monotonic
dependence of the rate of evolution on $s$ as implied by
Eq~(\ref{eq:v_mutation}), which is derived under the assumption that
$Ns\gg 1$.

Figure~\ref{fig:model1}A suggests that the rate at which
$\langle\phi\rangle$ increases at $t\simeq 4000$ displays non-monotonic
dependence on $s$.  In particular, the rate of evolution in
$\langle\phi\rangle$ seems to be maximized at $s\simeq 10^{-2}$.  To
confirm this result, we next computed the slope of $\left\langle \phi
\right\rangle$ (denoted by $v$) as a function of $s$. The value of $v$
was obtained by fitting a linear equation $vt+C$ to the curve of
$\left\langle \phi \right\rangle$ in the range of $t\in [2000, 4000]$ by
the least squares method. The result shows that $v$ depends
non-monotonically on $s$ (Fig~\ref{fig:model1}B), indicating that
evolution slows down as selection pressure increases, even though the
fitness landscape is smooth. The value of selection pressure maximizing
the rate of evolution is approximately $0.025$ for the parameters used
in Fig~\ref{fig:model1}.

The existence of a finite selection pressure maximizing the rate of
evolution $v$ stems from the two opposing effects of natural selection
on $v$. On the one hand, selection increases the fixation probability
of fitter genotypes, hence positively contributing to $v$. One the other
hand, selection decreases the residence time of immigrants, negatively
contributing to $v$, as described below. The genotypes of immigrants are
uncorrelated with those of the individuals already present in a
population. Thus, the immigrants can provide beneficial genes to the
population if they survive selection and recombine with the resident
individuals. However, the survival of the immigrants is hampered by
selection because their fitness is smaller than that of the resident
individuals. The duration for which the immigrants survive (the
residence time, for short) decreases with selection pressure. Therefore,
the probability that the population obtains beneficial genes through
recombination decreases as selection pressure increases. Owing to these
two opposing effects, there is a finite selection pressure maximizing
the rate of evolution.

The above intuitive argument can be made more quantitative by estimating
$v$ as a function of $s$ as follows. The value of $v$ is approximately
proportional to the rate at which novel beneficial genes are supplied to
a population (for simplicity, we ignore clonal interference between
resident individuals independently gaining beneficial genes from
migrants via recombination; this simplification is not expected to
affect our conclusion as described later). Such genes are supplied
through immigration followed by recombination between immigrants and
resident individuals. Thus, the rate of this supply is proportional to
$\mu$ (migration rate), $r$ (recombination rate), $N$ (population size),
$L$ (the number of loci), the residence time of migrants (denoted by
$\tau(s)$), and the probability that an immigrant carries a novel
beneficial gene per locus (denoted by $\rho$).  Furthermore, the
fixation of a beneficial gene occurs with a probability proportional to
$s$ for $s\ll 1$ and $Ns\gg 1$ \cite{Gil1998}.  Therefore, $v$ is
estimated as
\begin{eqnarray}
 v &\propto& N\rho rL \mu \tau(s) s
 \label{eq:v_recombination}
\end{eqnarray}
for $N^{-1}\ll s \ll 1$. Equation(\ref{eq:v_recombination}) differs from
Eq.~(\ref{eq:v_mutation}), in that the former contains $\tau(s)$, a
factor that negatively depends on $s$, whereas the latter contains no
such factor.

The probability $\rho$ generally depends on the fitness $\phi$; however,
$\rho$ can be regarded as constant in our simulations. Suppose that the
average number of beneficial genes in the genomes of resident
individuals is $l$. Then, a migrant has $l-\phi_0/2$ beneficial genes.
Recombination succeeds in increasing fitness only if a deleterious gene
of a resident individual is exchanged with a beneficial gene of a
migrant---this occurs with the probability $\rho=(L-l)/L\times
(l-\phi_0/2)/L$.  This probability $\rho$ takes the maximum at
$l=(L+\phi_0/2)/2\simeq L/2$.  Because we only consider a time range
within which $l\simeq L/2$ (more precisely, $500 \leq l < 550$ during
any simulation), $\rho$ can be regarded as nearly constant.

The residence time $\tau(s)$ can be roughly estimated as $e^{-\phi_0
  s}$, as follows.  First, we consider the situation in which one migrant
with fitness $e^{s\left( \phi - \phi_0 \right)}$ migrates into a
population of $(N-1)$ individuals with fitness $e^{s\phi}$.  The
probability that the migrant dies out in the next selection step is
calculated from Eq.~(\ref{eq:selection}) as
\begin{eqnarray}
 d &=& \left\{ \frac{(N-1)e^{s\phi}}{(N-1)e^{s\phi} + e^{s\left( \phi - \phi_0 \right)}} \right\}^N \simeq \exp{\left( -e^{-s\phi_0} \right)},
\end{eqnarray}
where we have used the fact that $N\gg 1$.  When we write the
probability distribution of residence time as $p(t)=(1-d)^td$, the
average residence time is calculated as
\begin{eqnarray}
 \left\langle t \right\rangle &=& \sum_{t=0}^\infty t p(t) = \frac{1-d}{d}.
\end{eqnarray}
By using the above expression for $d$, we finally obtain
\begin{eqnarray}
 \tau(s) &=& \left\langle t \right\rangle = \exp{\left( e^{-s\phi_0} \right)} -1 \simeq e^{-s\phi_0}
 \label{eq:tau_s}
\end{eqnarray}
for $s\phi_0\gg 1$.  

Taken together, the above results indicate that
\begin{eqnarray}
 v &\propto& N r \mu e^{-s\phi_0} s
 \label{eq:v_recombination_s}
\end{eqnarray}
for large values of $s\gg \phi_0^{-1}$. Therefore, $v$ decreases
exponentially for large $s$, whereas it increases linearly for small
$s$, with a crossover around $s_* \simeq 1/\phi_0$---i.e.\ $v$ depends
on $s$ non-monotonically. Equation~\ref{eq:v_recombination_s} also
implies that $v$ is proportional to $Nr\mu$. This implication is
supported by Fig~\ref{fig:model1}B, which shows that the values of $v$
for $Nr\mu=2\times 10^{-4}$ collapse into the same curve for different
values of $N$, $r$ and $\mu$, and that these values are almost twice the
values of $v$ for $Nr\mu=10^{-4}$, provided $s>0.1$. Moreover, a
semi-log plot of $v/(Nr\mu s)$ for various values of $N$, $r$, and $\mu$
shows that all data points collapse into a single line with a slope
close to $-\phi_0$ for $s>0.1$, as predicted by
Eq~\ref{eq:v_recombination_s} (Fig~\ref{fig:model1}D). Taken together,
these results support the validity of Eq~(\ref{eq:v_recombination_s}).

The derivation of Eq~(\ref{eq:v_recombination_s}) neglects clonal
interference between resident individuals that independently gain
beneficial genes from migrants. This simplification is unlikely to
affect the conclusion that $v$ depends on $s$ non-monotonically for the
following reason. The effect of clonal interference, which is always to
decrease $v$, is expected to diminish as $s$ increases, because the
residence time of a migrant decreases exponentially with $s$ according
to Eq~(\ref{eq:tau_s}). This expectation implies that allowing for
clonal interference would not alter non-monotonic dependence on $s$ itself but only shift the location of the maximum of $v$
in Eq~(\ref{eq:v_recombination_s}) along the $s$ axis. In addition, the
above expectation implies that clonal interference diminishes the
precision of Eq~(\ref{eq:v_recombination_s}) for small values of $s$, an
implication that might explain the dispersion of data points for $s<0.1$
in Fig~\ref{fig:model1}D.

The rate of evolution $v$ considered above is defined in terms of
changes in genotypes because $v$ is calculated from rescaled fitness
$\phi$. Alternatively, the rate of evolution can also be defined in
terms of changes in fitness. In this case, $sv$ rather than $v$ should
be considered because $sv$ is the rate of the change of $s\phi$, which
is a logarithm of fitness (i.e.\ Malthusian fitness). Under this
definition, there is still a finite selection pressure maximizing the
rate of evolution (Fig~\ref{fig:model1}C). Therefore, evolution can slow
down as selection pressure increases both in genotype space and in
fitness space (however, the latter result does not hold for Model 2 as
described below).

\section*{Model 2}
To investigate the generality of the results obtained with Model 1, we
next consider a situation in which multiple spatially-separate
subpopulations are evolving toward adaptation to a common ecological
niche. A subpopulation occasionally receives immigrants from the other
subpopulations. Since the subpopulations share the same niche, the fitness of
immigrants can be higher than that of resident individuals depending on
the degrees to which different subpopulations have adapted to the niche.
This situation may correspond to Model 1 with $\phi_0$ fluctuating
around $0$.  However, because determining the distribution of $\phi_0$
is difficult, here we explicitly consider multiple subpopulations.  As
in Model 1, mutation is ignored to focus on recombination-driven evolution.

\subsection*{Methods}
We consider a population consisting of $N_\mathrm{s}$ subpopulations.
Each subpopulation contains $N$ individuals.  The genotype of individual
$i\in \left\{ 1, \cdots, N \right\}$ in subpopulation $a\in \left\{ 1,
\cdots, N_\mathrm{s} \right\}$ is denoted by $\textrm{\boldmath
  $g$}_i^{(a)} \equiv \left( g_{1,i}^{(a)}, \cdots, g_{L,i}^{(a)}
\right)$.  Each variable $g_{l,i}^{(a)}$ takes the value $-1$
(deleterious) or $1$ (beneficial) as before.

The time evolution of the system consists of three steps as in Model 1:
selection, recombination, and migration.  The selection step in each subpopulation is the same
as in Model 1 (see Eqs.~(\ref{eq:fitness}) and (\ref{eq:selection})). In
the recombination step, individuals in the same subpopulation exchange
genes.  The exchange
\begin{eqnarray}
 && \left( \cdots, g_{l-1,i}^{(a)}, g_{l,i}^{(a)}, g_{l+1,i}^{(a)}, \cdots \right) + \left( \cdots, g_{l-1,j}^{(a)}, g_{l,j}^{(a)}, g_{l+1,j}^{(a)}, \cdots \right) \nonumber \\
 &\rightarrow& \left( \cdots, g_{l-1,i}^{(a)}, g_{l,j}^{(a)}, g_{l+1,i}^{(a)}, \cdots \right) + \left( \cdots, g_{l-1,j}^{(a)}, g_{l,i}^{(a)}, g_{l+1,j}^{(a)}, \cdots \right)
\end{eqnarray}
occurs with probability $r$ per individual per locus per
generation. Pairs of individuals undergoing recombination are selected
randomly.

In the migration step, individuals migrate between subpopulations.
Individuals change as
\begin{eqnarray}
 \textrm{\boldmath $g$}_i^{(a)} &\rightarrow& \textrm{\boldmath $g$}_j^{(b)}
 \label{eq:migration1}
\end{eqnarray}
for each pair of individuals $(i,j)$ and for each pair of subpopulations
$(a,b)$ with probability $D$ (spatial structure is ignored).  That is,
individual $i$ in subpopulation $a$ is replaced by a copy of the
individual $j$ in subpopulation $b$.

We set initial conditions as random configurations with rescaled fitness
$\phi=0$ for each genotype $\textrm{\boldmath $g$}_i^{(a)}$.  Parameters
were set as $N=1000$, $N_\mathrm{s}=64$, $L=1000$, $r=10^{-4}$, and
$D=10^{-7}/64^2$.  Statistical quantities were calculated by running
$1000$ replicate simulations.


\subsection*{Results}
\begin{figure}[!h]
\begin{adjustwidth}{-2.25in}{0in}
\includegraphics[clip]{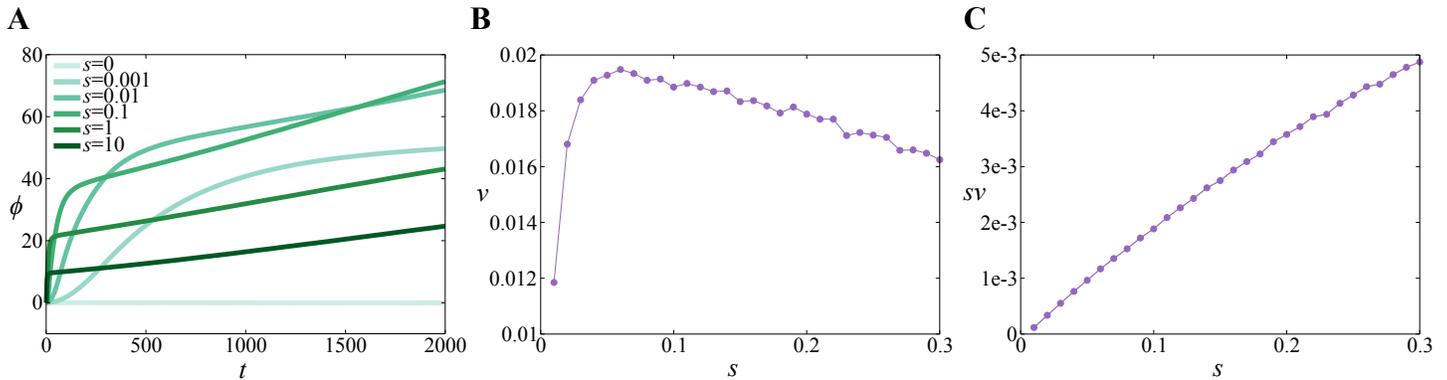}
\caption{\textbf{A.} The average rescaled fitness $\left\langle \phi
  \right\rangle$ as a function of time $t$ for various strengths of
  selection (denoted by $s$) for Model 2. $\phi$ is proportional to the
  number of beneficial alleles in a genome (see Table~\ref{tbl:notation}
  for details). \textbf{B.}  The rate of evolution $\Delta \phi/\Delta
  t$ (denoted by $v$) as a function of $s$ for Model 2. \textbf{C.} The
  rate of evolution in Malthusian fitness (i.e.\ a logarithm of fitness)
  $sv$ as a function of $s$ for Model 2. }
\label{fig:model2}
\end{adjustwidth}
\end{figure}

We display the time evolution of the average rescaled fitness
$\left\langle \phi \right\rangle$ in Fig~\ref{fig:model2}A.  We find
that the dynamics of $\left\langle \phi \right\rangle$ consists of the
two phases, as in Model 1.  In addition to saturation due to the
finiteness of $L$, it should be noted that $\left\langle \phi
\right\rangle$ saturates for $t\gg 1000$ because genotypes of all
subpopulations eventually become homogeneous in recombination-driven
evolution.  However, this saturation is expected to disappear as
$N_\mathrm{s}$ increases to infinity, and we focus on the second phase
driven by both recombination and migration.  The slope of the linear
parts in $t\simeq 2000$ has non-monotonic $s$ dependence, a result that
is the same as in Model 1. The slope $v$ of this linear region is also
estimated by fitting of the $t\in [1000,2000]$ part of the curves, and is
plotted for various $s$ in Fig~\ref{fig:model2}B.  We observe the
existence of the finite selection pressure value maximizing the rate of
evolution at $s\simeq 0.06$.

The mechanism of this phenomenon is explained as follows.  In the second
phase, a quasi-steady state is realized and dominant process is
migration.  When an individual migrates into a subpopulation, the
migrant and other individuals in the subpopulation can incorporate
the beneficial genes of each other by recombination.
This process most likely succeeds in intermediate selection pressure $s$.  For
small $s$, selection does not work effectively.  For large $s$, the
subpopulation does not have enough time to take in the information
provided by the migrant.  Therefore, there exists finite selection
pressure maximizing the rate of evolution.

For Model 1, we found that the selection pressure maximizing the
velocity in fitness space $sv$ is also finite as displayed in
Fig~\ref{fig:model1}C.  In contrast, we find that a selection pressure
maximizing $sv$ is infinite for Model 2, as displayed in
Fig~\ref{fig:model2}C.  We note that Eq~(\ref{eq:v_recombination}) is
also expected to hold for Model 2.  We therefore consider that the
difference between the two models comes from the difference in
the meaning of $\tau(s)$.  In Model 1, migration always decreases the
fitness by definition, and $\tau(s)$ is the residence time of the
migrant before dying out.  In Model 2, migration can both increase and
decrease the fitness of a subpopulation, and $\tau(s)$ is the time
during which the migrant stays in the subpopulation.  Since migrants
with higher fitness do not necessarily die out in Model 2, $\tau(s)$ does not
decrease exponentially for large $s$, but decreases much slowly.  This
slower decrease in $\tau(s)$ is likely to be the reason why $sv$ does not have a
maximum at finite $s$, although $v$ has.  Further investigation will be
needed for this topic.


\section*{Discussion}
In this paper, we investigated evolution
driven by selection, recombination and migration on a smooth fitness landscape.  We find that the
speed of evolution can slow down as selection pressure $s$ increases
without the ruggedness of the fitness landscape $\phi\left(
\textrm{\boldmath $g$} \right)$.  Our results suggest that an optimal
selection pressure exists for evolution driven by recombination, in
contrast to evolution driven by mutation.

Before ending this paper, we make six remarks.  The first remark is
related to the inclusion of the effect of mutation.  We conjecture that
whether mutation or recombination is dominant is determined by comparing two
expressions (\ref{eq:v_mutation}) and (\ref{eq:v_recombination}).  The
selection pressure maximizing the rate of evolution will become infinity
in the presence of frequent mutation in addition to recombination, while
our recombination-driven result will be reproduced in a weak mutation
regime.  At the simplest level, we estimate that mutation-driven
situation is realized when $u\gg r \mu \tau(s)$, while
recombination-driven situation is realized when $u\ll r \mu \tau(s)$.
The verification of this conjecture and analysis of the intermediate
regime $u\simeq r \mu \tau(s)$ are one of the future problems.

The second remark is related to the biological relevance of our
results.  In our study, we have focused on recombination-driven situation and ignored mutation. In evolution of
prokaryotes, there are many situations where recombination rate is much
larger than mutation rate.  In fact, we can see spontaneous horizontal
gene transfer rate in some prokaryotes is larger than mutation rate in
\cite{Oveet2013}.  Furthermore, there is another evidence that evolution
is mainly driven by horizontal gene transfer in some prokaryotes, as
reported in \cite{FSES2000, VosDid2009, Puiet2014, LinKus2017}. 
Since we consider the situation where selection is stronger than recombination, the result is applicable mainly to prokaryotes, and the applicability to sexually-reproducing eukaryotes with larger recombination rate may be limited.
Note that mutation in realistic
systems is almost deleterious, and the rate of mutation flipping genes
as in our model is expected to be further small. Therefore, when
population size is relatively small, a phenomenon reported in this paper
may be observed in some prokaryotes.

Third, our study might look similar to a study by Barton \cite{Bar1983}, in that both
investigate the dynamics of introgression in the presence of linkage
disequilibrium. However, our study differs from that of Barton in the
following aspect: whereas Barton's study considers the introgression of
deleterious alleles linked to each other, our study considers
the introgression of beneficial alleles linked to deleterious alleles.
Our study also differs in the conclusion about how introgression depends on
the strength of selection. Barton's study shows that the introgression
of deleterious alleles is a monotonic function of the selection coefficient. In
contrast, our study shows that the introgression of beneficial alleles is a
non-monotonic function of the selection coefficient, the result
essentially due to the interactions between beneficial and deleterious
alleles through linkage disequilibrium.

Fourth, previous studies find that an optimal recombination rate exists
in evolution driven by mutation and recombination \cite{CKL2005,
  LWK2016}.  One might think that our result on the optimal value in
selection pressure can be trivially deduced from their results.
However, we believe that this is not the case because in our case the
optimal value of $s$ is almost independent of $r$, as seen from
Eq~(\ref{eq:v_recombination}).  In addition, an optimal recombination
rate in \cite{CKL2005} may come from the loss of genetic diversity due
to copy-and-paste-type recombination, which differs from recombination
considered in our models (see Eq~(\ref{eq:recombination})).  The paper
\cite{LWK2016} reports an optimal recombination rate in a state in which
fitness is stationary over time, whereas we here focus on a state in
which fitness steadily increases over time.  Therefore, we think that
the mechanism by which an optimal $s$ value arises investigated in our
work differs from those by which optimal recombination rates arise
investigated in the previous studies.

Fifth, evolution can slow down even in a mutation-driven situation when
a fitness landscape is rugged and population is finite; however, this
differs from our result.  When selection pressure is too strong, whole
population gets stuck into a local maximum of a rugged fitness
landscape, and the speed of evolution becomes small.  Although this
mechanism has not been studied systematically, the population-size
dependence of the speed of evolution on rugged fitness landscapes has
recently attracted much attention \cite{RHHV2008, HanRoz2009, JKP2011,
  OchDes2015}.  However, whereas these phenomena result from the
ruggedness of fitness landscapes, those reported in this paper come
from the decrease of genetic variation due to selection. Therefore, we
believe that the two are different phenomena.

We finally remark that the phenomenon reported in this paper may be
similar to negative differential resistance (NDR) \cite{SGN1997,
  ZPM2002, KMHLT2006, Sel2008, BBM2013, BIOSV2014, BSV2015}.  NDR is a
phenomenon in which particle current becomes smaller by increasing
external force.  NDR has been observed in many physical systems, and is
regarded as a common property of transport in crowded environments.  In
our paper, we find a phenomenon in which the rate of evolution becomes
smaller by increasing selection pressure.  The fact that recombination
does not work as population becomes homogeneous seems to be similar to
the fact that particles cannot move as the positions of particles become
close to each other in kinetically constrained models (KCM)
\cite{RitSol2003}, which are one of the models for glass with smooth energy
landscapes and exhibit NDR.  In this analogy, selection pressure
corresponds to external force for KCM \cite{Sel2008}, and the fact that
fitness cannot increase as genotypes become homogeneous corresponds to
the fact the particles cannot flow in the direction of external force as
particles get crowded.  Therefore, while rugged fitness
landscape models are similar to spin glass models \cite{MPV1987}, our
model may be similar to KCM.  Further similarity to KCM will be studied
in future.

\section*{Acknowledgments}
We thank Nen Saito for fruitful discussions. The present study was
supported by JSPS KAKENHI Grant Numbers JP15H05746, JP16J00178, JP17K17657,
and by the Platform for Dynamic Approaches to Living System
from Japan Agency for Medical Research and Development (AMED).


%
%
%

\end{document}